\documentclass[floatfix,epsfig,twocolumn,amsmath,amssymb]{revtex4}
\usepackage{epsfig}

\def\snn{\sqrt{s_{_{\rm NN}}}}
\def\PHOBOS{P\kern-.34em \lower.4ex\hbox{H}\kern-.12em \lower.6ex\hbox{O}\kern-.08em \lower-.18ex\hbox{B}\kern-.12em \lower-.45ex\hbox{O}\kern-.12em \lower.12ex\hbox{S}\ }
\def\FigureOne{
  \begin{figure}{}
    \centerline{ \epsfig{file=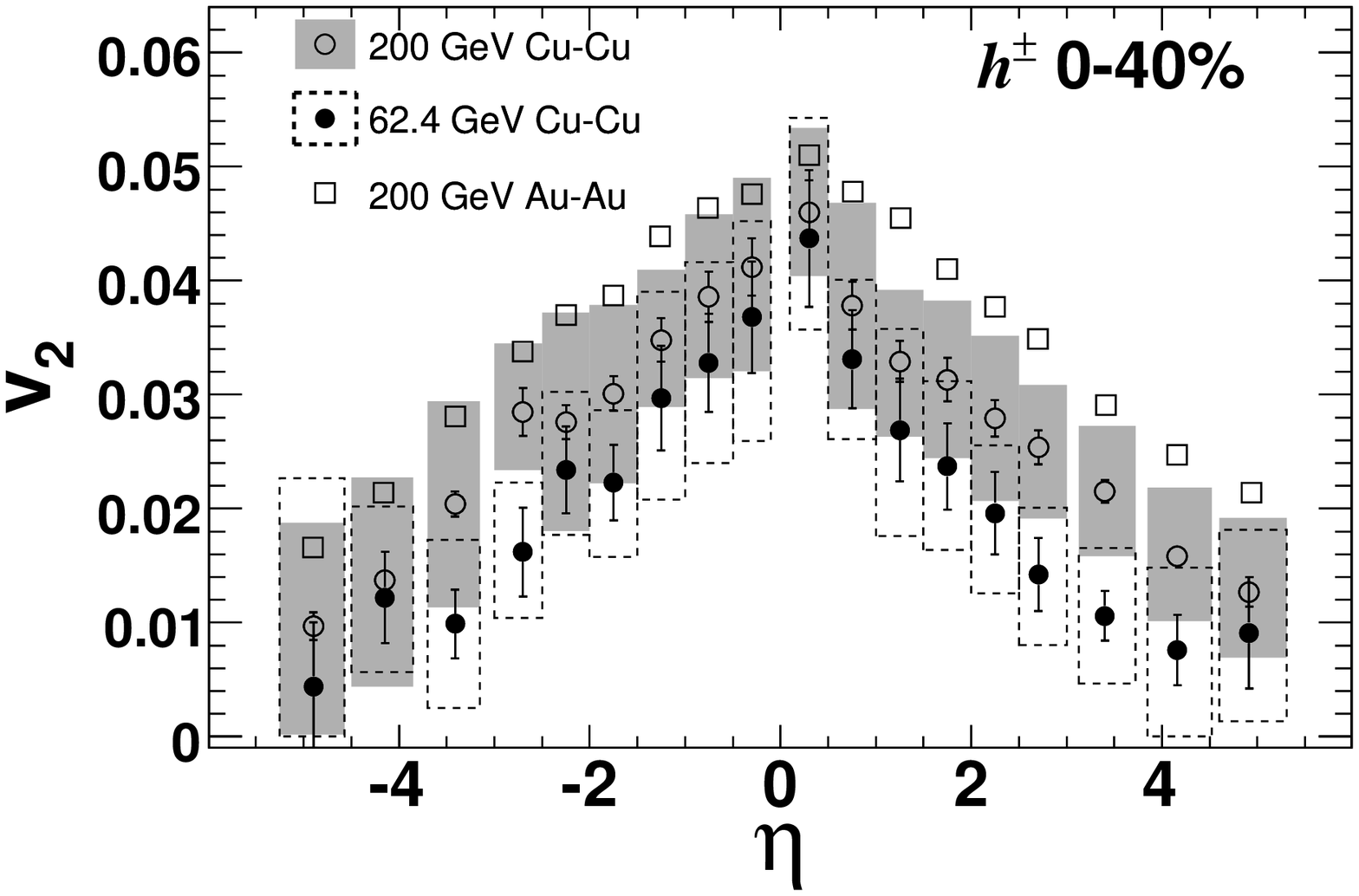,width=9.0cm}}
    \caption{v$_2$ vs.\ $\eta$ for Cu-Cu collisions at
      $\snn =$ 62.4 and 200 GeV using the hit-based analysis.  The boxes
      show the 90\% C.L.\ systematic errors and the
      bars represent the 1-$\sigma$ statistical errors.
      Previously published 200 GeV Au-Au data (without error bars)
      is shown for comparison.}
    \label{v2etaCu}
  \end{figure} }
\def\FigureTwo{
  \begin{figure}{}
    \centerline{ \epsfig{file=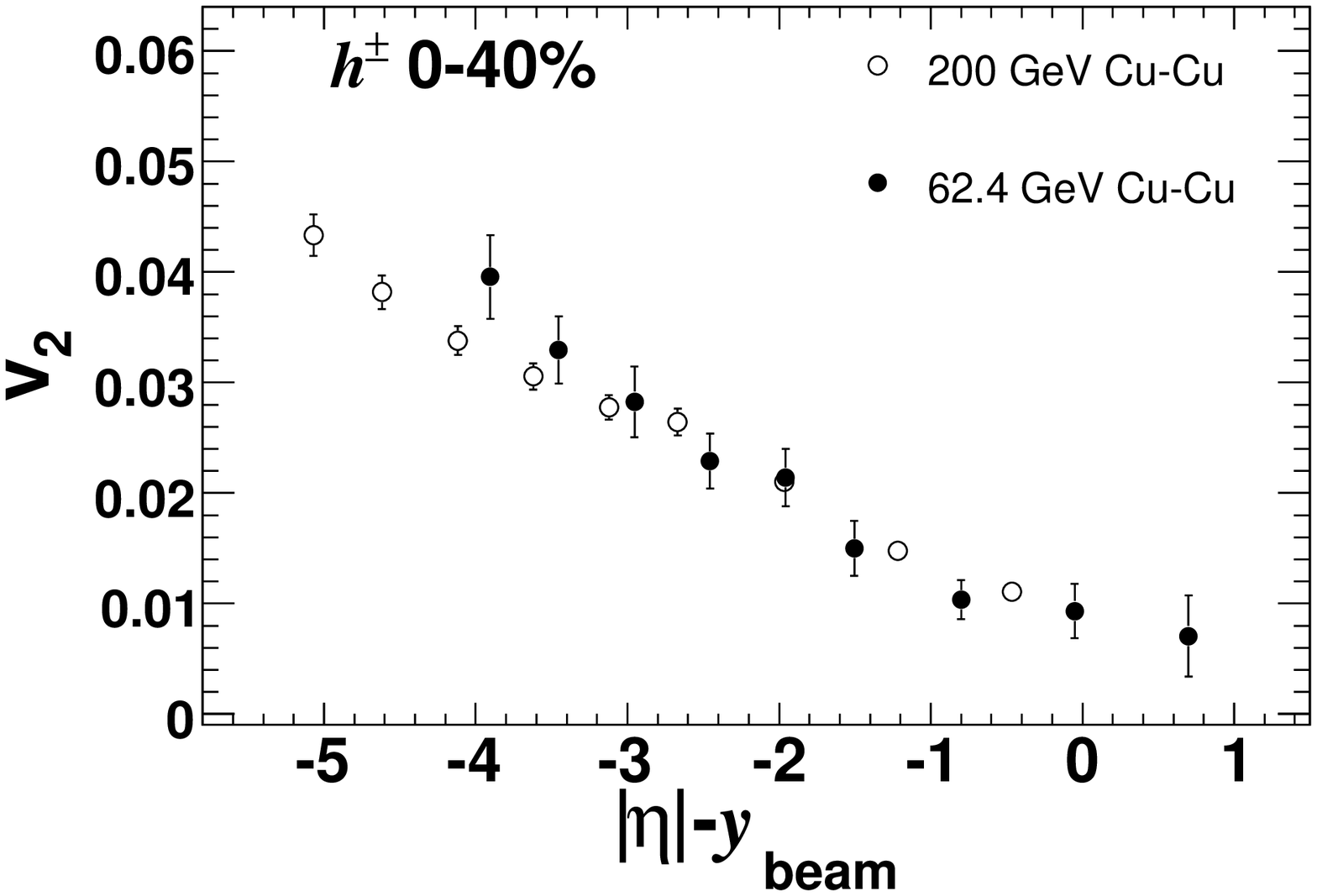,width=9.0cm}}
    \caption{v$_2$ vs.\ $|\eta|-$y$_{\rm beam}$ for Cu-Cu collisions at
      $\snn =$ 62.4 and 200 GeV from the hit-based analysis.
      Only 1-$\sigma$ statistical errors are shown.}
    \label{v2etapCu}
  \end{figure} }
\def\FigureThree{
  \begin{figure}{}
    \centerline{ \epsfig{file=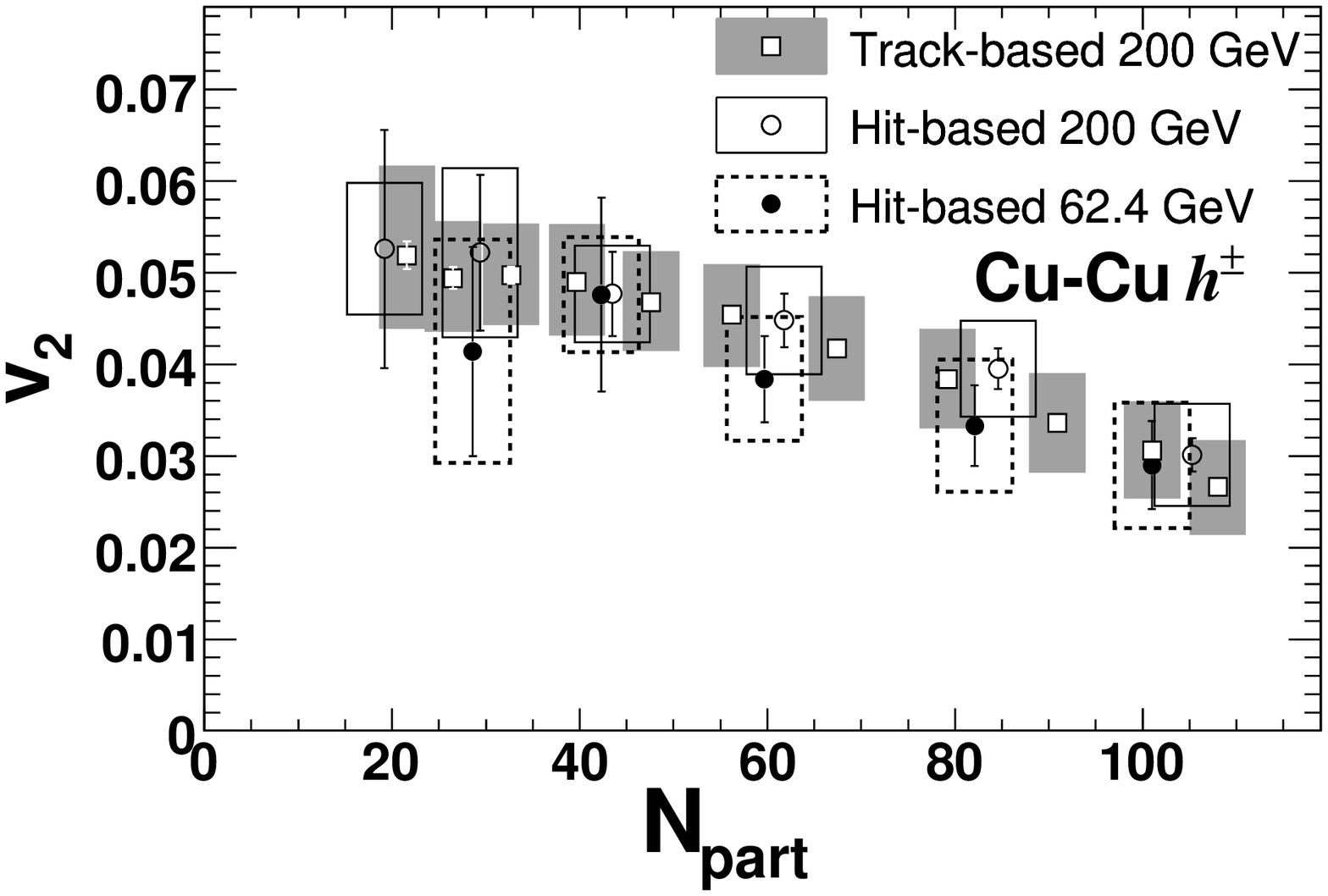,width=9.0cm}}
    \caption{v$_2$ vs.\ N$_{\rm part}$ for Cu-Cu collisions at
      $\snn =$ 62.4 and 200 GeV.  The boxes
      show the 90\% C.L.\ systematic errors and the
      lines represent the 1-$\sigma$ statistical errors.
      The results from two analysis methods are
      shown, as discussed in the text. v$_2$ is shown for $|\eta| < 1$
      and $0 < \eta < 1$ for the hit-based and track-based methods,
      respectively.}
    \label{v2npartCu}
  \end{figure} }
\def\FigureFour{
  \begin{figure}{}
    \centerline{ \epsfig{file=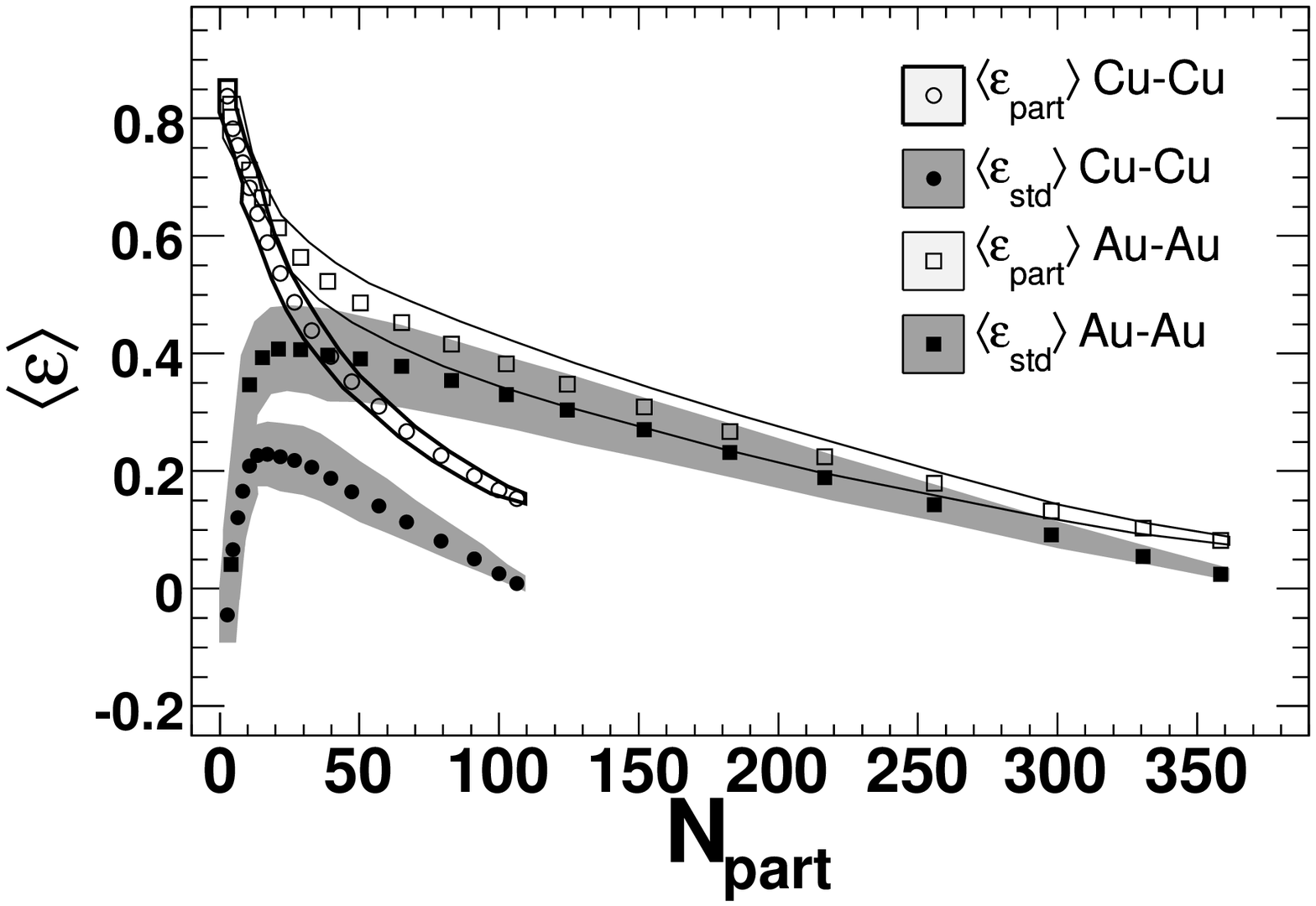,width=9.0cm}}
    \caption{The average eccentricity defined in two ways,
      ($\langle \varepsilon_{\rm std} \rangle$ and
      $\langle \varepsilon_{\rm part} \rangle$), as described
      in the text, vs.\ N$_{\rm part}$ for simulated Cu-Cu and Au-Au
      collisions at $\snn =$ 200 GeV.
      The bands show the 90\% C.L.\ systematic errors.}
    \label{aveecc}
  \end{figure} }
\def\FigureFive{
  \begin{figure}{}
    \centerline{ \epsfig{file=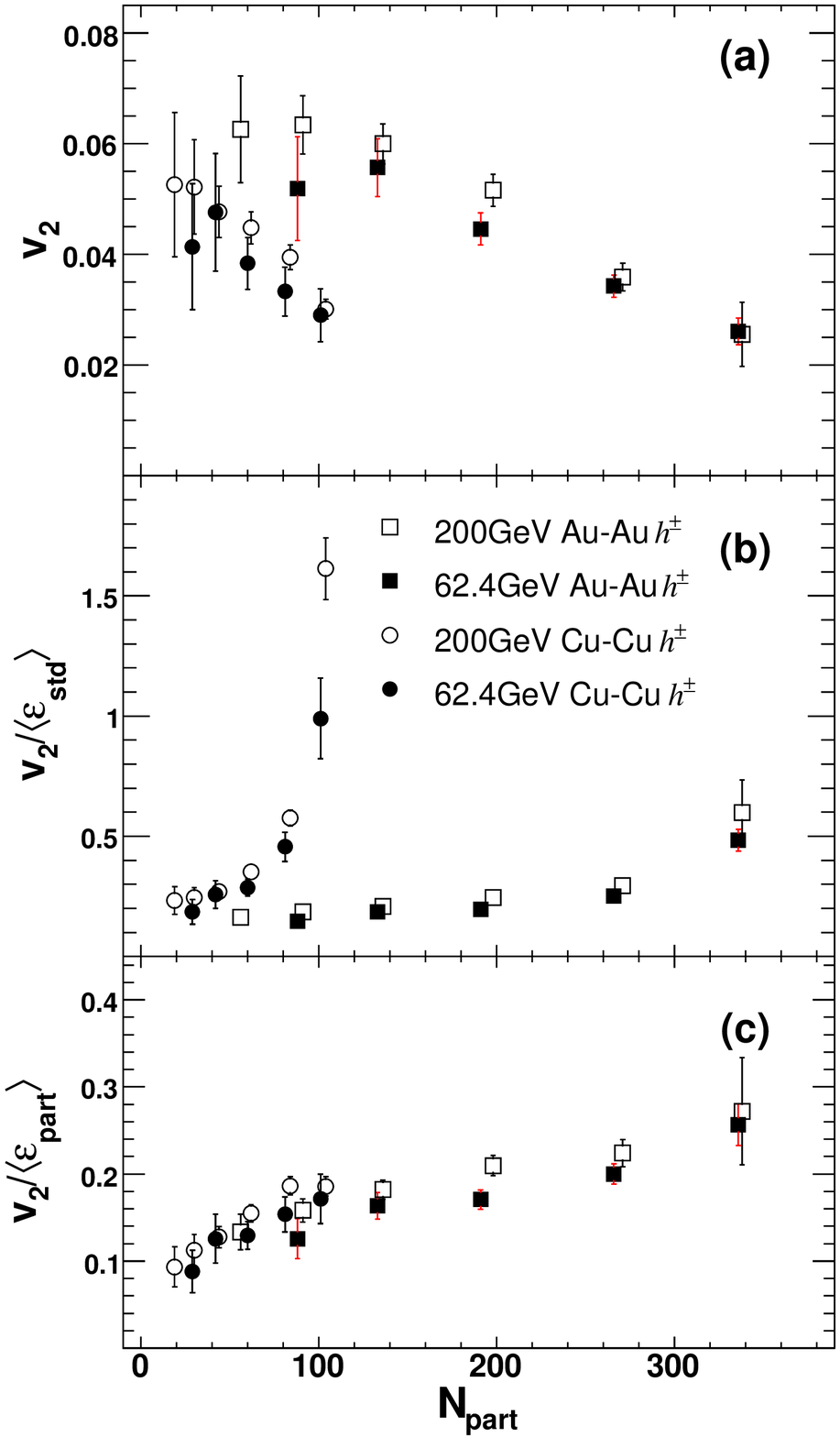,width=9.0cm}}
    \caption{(a) v$_2$ (unscaled) vs.\ N$_{\rm part}$,
      (b) v$_2$/$\langle \varepsilon_{\rm std} \rangle$
      vs.\ N$_{\rm part}$, and (c) v$_2$/$\langle \varepsilon_{\rm part}
      \rangle$ vs.\ N$_{\rm part}$, for
      Cu-Cu and Au-Au collisions at $\snn =$ 62.4 and 200 GeV.
      1-$\sigma$ statistical error bars are shown.
      v$_2$ is shown in $|\eta| < 1$ for the hit-based method.}
    \label{v2nparteccscale}
  \end{figure} }

\begin{document}

\title{System Size, Energy, Pseudorapidity, and Centrality Dependence \\
    of Elliptic Flow}

\author{
B.Alver$^4$,
B.B.Back$^1$,
M.D.Baker$^2$,
M.Ballintijn$^4$,
D.S.Barton$^2$,
R.R.Betts$^6$,
A.A.Bickley$^7$,
R.Bindel$^7$,
W.Busza$^4$,
A.Carroll$^2$,
Z.Chai$^2$,
V.Chetluru$^6$,
M.P.Decowski$^4$,
E.Garc\'{\i}a$^6$,
T.Gburek$^3$,
N.George$^2$,
K.Gulbrandsen$^4$,
C.Halliwell$^6$,
J.Hamblen$^8$,
I.Harnarine$^6$,
M.Hauer$^2$,
C.Henderson$^4$,
D.J.Hofman$^6$,
R.S.Hollis$^6$,
R.Ho\l y\'{n}ski$^3$,
B.Holzman$^2$,
A.Iordanova$^6$,
E.Johnson$^8$,
J.L.Kane$^4$,
N.Khan$^8$,
P.Kulinich$^4$,
C.M.Kuo$^5$,
W.Li$^4$,
W.T.Lin$^5$,
C.Loizides$^4$,
S.Manly$^8$,
A.C.Mignerey$^7$,
R.Nouicer$^2$,
A.Olszewski$^3$,
R.Pak$^2$,
C.Reed$^4$,
E.Richardson$^7$,
C.Roland$^4$,
G.Roland$^4$,
J.Sagerer$^6$,
H.Seals$^2$,
I.Sedykh$^2$,
C.E.Smith$^6$,
M.A.Stankiewicz$^2$,
P.Steinberg$^2$,
G.S.F.Stephans$^4$,
A.Sukhanov$^2$,
A.Szostak$^2$,
M.B.Tonjes$^7$,
A.Trzupek$^3$,
C.Vale$^4$,
G.J.van~Nieuwenhuizen$^4$,
S.S.Vaurynovich$^4$,
R.Verdier$^4$,
G.I.Veres$^4$,
P.Walters$^8$,
E.Wenger$^4$,
D.Willhelm$^7$,
F.L.H.Wolfs$^8$,
B.Wosiek$^3$,
K.Wo\'{z}niak$^3$,
S.Wyngaardt$^2$,
B.Wys\l ouch$^4$\\
\vspace{3mm}
\small
$^1$~Physics Division, Argonne National Laboratory, Argonne, IL 60439-4843,
USA\\
$^2$~Chemistry and C-A Departments, Brookhaven National Laboratory, Upton, NY
11973-5000, USA\\
$^3$~Institute of Nuclear Physics PAN, Krak\'{o}w, Poland\\
$^4$~Laboratory for Nuclear Science, Massachusetts Institute of Technology,
Cambridge, MA 02139-4307, USA\\
$^5$~Department of Physics, National Central University, Chung-Li, Taiwan\\
$^6$~Department of Physics, University of Illinois at Chicago, Chicago, IL
60607-7059, USA\\
$^7$~Department of Chemistry, University of Maryland, College Park, MD 20742,
USA\\
$^8$~Department of Physics and Astronomy, University of Rochester, Rochester,
NY 14627, USA\\ }
\date{\today}

\begin{abstract} \noindent
This paper presents measurements of the elliptic flow of charged
particles as a function of pseudorapidity and centrality from Cu-Cu
collisions at 62.4 and 200 GeV using the PHOBOS detector at the Relativistic
Heavy Ion Collider (RHIC). The elliptic flow in Cu-Cu collisions is
found to be significant even for the most central events.
For comparison with the Au-Au results, it is found that the detailed way
in which the collision geometry (eccentricity) is estimated is of critical
importance when scaling out system-size effects.
A new form of eccentricity, called the {\sl participant
eccentricity}, is introduced which yields a scaled elliptic flow
in the Cu-Cu system that has the same relative magnitude and
qualitative features as that in the Au-Au system.
\end{abstract}

\maketitle

The characterization of the collective flow of produced particles by their
azimuthal anisotropy has proven to be one of the more fruitful probes of
the dynamics of heavy ion collisions at the Relativistic Heavy Ion Collider
(RHIC). Flow is sensitive to the early stages of the collision and so the
study of flow affords unique insights into the properties of the hot,
dense matter that is produced, including information about the degree of
thermalization and its equation of state~\cite{Kolb2000}. In particular,
flow measurements as a function of pseudorapidity have motivated the
development of three-dimensional hydrodynamic models of relativistic
heavy ion collisions~\cite{Hirano} and provide information crucial in
constraining these models and others that seek to better understand what
role the longitudinal dimension plays in the collision%
~\cite{HeinzKolb,Buda-Lund}.
By examining the successes and failures of hydrodynamic models of heavy ion
collisions, a better understanding of the dynamics of these collisions
is gained.

Elliptic flow has been studied extensively in Au-Au collisions at RHIC as
a function of pseudorapidity ($\eta$), centrality, transverse momentum,
and energy~\cite{phflow,NARC,WP,STARWP,PhenWP}. The large
pseudorapidity coverage of the PHOBOS detector makes it ideally suited for
probing the longitudinal structure of the collision, the dynamics of
which have only recently begun to be understood away from
midrapidity~\cite{DisEff}.
This work presents new results on data taken by the PHOBOS
experiment at RHIC showing a detailed comparison of differential
measurements of elliptic flow in Cu-Cu and Au-Au collisions
at $\snn =$ 62.4 and 200 GeV.

The strength of the elliptic flow, v$_2$, is given by the coefficient
of the second harmonic in the Fourier expansion of the particle
azimuthal angle distribution relative to the reaction plane,
$\Psi_{\rm R}$. For this analysis $\Psi_2$, an estimate of
$\Psi_{\rm R}$, was used as in Ref.~\cite{PandV}.

The PHOBOS detector is comprised of silicon pad detectors for
tracking, vertex detection, and multiplicity measurements. Details
of the setup and the layout of the silicon sensors can be found
elsewhere~\cite{PhobosDet}. Key elements of the detector used in this
analysis include the first six layers of both silicon spectrometer
arms, the silicon vertex detector (VTX), the silicon octagonal
multiplicity detector (OCT), three annular silicon multiplicity
detectors to either side of the collision point, and two sets of
scintillating paddle counters for centrality determination.

Monte Carlo simulations of the detector performance were based on
the HIJING event generator~\cite{HIJING} and the
GEANT~3.211~\cite{GEANT} simulation package, folding in the signal
response for scintillator counters and silicon sensors.

The data shown here were taken with the PHOBOS detector at RHIC during
the years 2001--2005.  The Au-Au data are published in 
previous papers describing work at $\snn =$ 19.6, 62.4, 130,
and 200 GeV~\cite{phflow,limfrag,NARC}. The Cu-Cu data at $\snn =$
62.4 and 200 GeV presented here were analyzed in a similar fashion,
using the hit-based method that utilized hits in the VTX, OCT and
ring sub-detectors to measure flow over a wide range in pseudorapidity
($|\eta| < 5.4$) and a track-based method that made use of tracks in
the spectrometer arms and had a smaller pseudorapidity coverage
($0.0 < \eta < 1.0$). For details on the hit-based and track-based
methods, see Refs.~\cite{phflow} and \cite{NARC}, respectively.

The event-by-event collision vertex was determined using the intersection of 
tracks identified in the spectrometer and extrapolated back 
to a common point.  
The flow analysis was based on the anisotropy of the azimuthal 
distribution of
charged particles traversing the detector. At the points where
charged tracks passed through an active silicon detector, energy
was deposited in the form of ionization. A pad where energy
was deposited is said to be a ``hit''.
This analysis is based on the ``sub-event'' technique wherein%
\FigureOne%
one studies the correlation of hits in one part of the detector
with the event plane angle as determined by hits in a different
part of the detector~\cite{PandV}.  As described in the earlier work~%
\cite{phflow,NARC,limfrag,v1limfrag}, 
corrections are applied to account for signal dilution due to 
detector occupancy and adjustments are made to create an appropriately 
symmetric acceptance for the analysis.  The sub-event regions used in the 
event plane calculation were $0.1<|\eta|<3.0$ for both 62.4 and 200 GeV. 
The event plane resolution was calculated separately for each
centrality bin. The resolution correction factor ranged from 2 to 3 on
average, with the larger correction necessary at 62.4 GeV. For the
determination of v$_2$ in the positive (negative) $\eta$ region
of the detector, the sub-event on the opposite side of midrapidity
was used to evaluate $\Psi_2$.

Monte Carlo simulations showed a residual suppression of the flow
signal dominated by background particles carrying
no flow information and the loss of sensitivity due to the hit map
symmetrization and the occupancy correction algorithm. As in our
earlier work with the hit-based technique, this suppression was
corrected using simulated data
by comparing the output resolution corrected flow signal to the
input flow signal for many samples of simulated data with
different shapes and magnitudes of input flow signal.

Numerous sources of systematic error were investigated, including
effects due to the hit definition, hit merging, sub-event
definition, knowledge of the beam orbit relative to the detector,
hole filling procedure, vertexing algorithm, and suppression
correction determination. The effect of these sources depended
both on $\eta$ and centrality. In general, the systematic error
arising from each source was determined by varying that specific
aspect of the analysis (or several aspects in concert) within
reasonable limits and quantifying the change in the final
v$_2$ result as a function of $\eta$ and centrality. The
individual contributions were added in quadrature to derive
the 90\% confidence level error shown in the results presented
here.  The systematic uncertainty was dominated by the
suppression correction determination.

Figure~\ref{v2etaCu} shows the elliptic flow signal as a
function of pseudorapidity in Cu-Cu collisions at $\snn =$ 62.4
and 200 GeV for the 40\% most central events.  The
\FigureTwo
resemblance to published Au-Au results~\cite{limfrag} (also shown in
Figure~\ref{v2etaCu}) is striking. The Cu-Cu v$_2$ displays a similar
shape in pseudorapidity to that of Au-Au, with a magnitude at midrapidity
only 10--20\% lower than that seen in Au-Au, increasing to $\sim$40\%
at large $|\eta|$. The strength of the Cu-Cu v$_2$
signal is surprising in light of expectations that the smaller system
size would result in a much smaller flow signal~\cite{ChenKo}.

The Cu-Cu v$_2$ also exhibits extended longitudinal scaling, as shown in
Figure~\ref{v2etapCu}, and as already seen in Au-Au collisions
for elliptic flow~\cite{limfrag} and directed flow~\cite{v1limfrag}
and for charged particle multiplicity~\cite{BRAHMS,PMLF,multlimfrag}. The
agreement between the two energies in $|\eta|-y_{\rm beam}$ implies
that, as with Au-Au, the elliptic flow is largely independent of
energy when viewed (effectively) in the rest frame of one of the
colliding nuclei.

The centrality dependence of the elliptic flow measured in Cu-Cu is
presented in Figure~\ref{v2npartCu}, where v$_2$ is plotted as a function
of the number of participating nucleons, N$_{\rm part}$. Both hit-based
and track-based analyses were used for the 200 GeV data, and the results
of the two methods agree quite well within errors.

A substantial flow signal is measured in Cu-Cu at both energies for even%
\FigureThree%
the most central events. This is quite surprising, as the initial
spatial anisotropy gives rise to a momentum space anisotropy
which, in turn, produces the flow~\cite{flowprod}. It is expected
therefore that v$_2$ should approach zero as the collisions become
more central, as it does
for Au-Au~\cite{NARC}. The persistent and non-trivial elliptic
flow signal seen in the most central events implies that something
beyond the expected nuclear collision geometry may be responsible 
for driving the flow signal.  

To explore this question further, it is 
useful to compare directly the elliptic flow signal across 
different colliding species, {\sl i.e.}, make a direct comparison between 
the flow seen in Cu-Cu and Au-Au collisions. To do this, it is
important to scale out the difference in the initial geometric
asymmetry of the collision, {\sl i.e.}, the eccentricity of the
collision. This is crucial since in each selected centrality range
the average eccentricity depends on the size of the colliding
species.

Typically, the eccentricity is defined by relating the
impact parameter of the collision in a Glauber model simulation to
the eccentricity calculated assuming the minor axis of the overlap
ellipse to be along the impact parameter vector. Thus, if the
$x$-axis is defined to be along the impact parameter vector and
the $y$-axis perpendicular to that in the transverse plane, the
eccentricity is determined by~\cite{Heiselberg,Sorge}
\begin{equation}
\varepsilon=\frac{\sigma_{y}^{2}-\sigma_{x}^{2}}
{\sigma_{y}^{2}+\sigma_{x}^{2}},
\end{equation}
where $\sigma_{x}$ and $\sigma_{y}$ are the RMS widths of the
participant nucleon distributions projected on the $x$- and $y$-axes,
respectively. Let us call the eccentricity determined in
this fashion $\varepsilon_{\rm std}$.

\FigureFour
The relation of the eccentricity to the centrality depends on the 
details of the eccentricity definition used in the Glauber model simulation.  
The definition most commonly used is presented above.  Implicit in this 
choice is a physics bias about the relevant asymmetry that drives 
the flow signal. It is important to consider other possibilities.  
In particular, a natural choice to consider is the geometry
of the participant nucleons themselves.  

\FigureFive
In a large system, the nuclear 
geometry and the participant geometry largely coincide. For small 
systems or small transverse overlap regions, however, fluctuations
in the nucleon positions in Glauber model calculations, as described below,
frequently create a
situation where the minor axis of the ellipse in the transverse
plane formed by the participating nucleons is not along the impact
parameter vector. One way to address this issue is to make a
principal axis transformation, rotating the $x$- and $y$-axes used
in the eccentricity definition in the transverse plane in such a
way that $\sigma_{x}$ is minimized. Let us call the eccentricity
determined in this fashion the participant eccentricity,
$\varepsilon_{\rm part}$, and the plane specified by the beam axis and the
$x$-axis in the participant frame $\Psi_{\rm participant}$. In terms of the original $x$- and $y$-axes
(in fact, any pair of perpendicular transverse axes),
\begin{equation}
\varepsilon_{\rm part}=\frac{
\sqrt{(\sigma_{y}^{2}-\sigma_{x}^{2})^{2}+4(\sigma_{xy})^{2}} }
{\sigma_{y}^{2}+\sigma_{x}^{2}}.
\end{equation}
%
In this formula, $\sigma_{xy}=\langle xy \rangle - \langle x
\rangle \langle y \rangle$. The average values of
$\varepsilon_{\rm std}$ and $\varepsilon_{\rm part}$ are
quite similar for all but the most peripheral interactions for
large species, as is shown in Figure~\ref{aveecc} for Au-Au.  For
smaller systems such as Cu-Cu, however, fluctuations in the nucleon
positions in Glauber model calculations (described below) become quite
important for all centralities and the
average eccentricity can vary significantly depending on how it is
calculated.  This is also illustrated in Figure~\ref{aveecc}.

The effects of finite number fluctuations on elliptic flow have been studied
for large collision systems with Monte-Carlo
simulations~\cite{Ollitrault,Miller,Zhu} and were found to be
small.  However, in Cu-Cu collisions these fluctuations
are larger and could have a significant impact on the elliptic flow.  The
participant eccentricity allows these fluctuating configurations to be
considered seriously on an event-by-event basis.

The Glauber model used for the calculation of these eccentricities
is a Monte Carlo toy model which builds nuclei by randomly placing nucleons
according to a Woods-Saxon distribution.  Excluded volume effects
were incorporated into the model, requiring a minimum center-to-center
nucleon separation of 0.4 fm, to agree with HIJING~\cite{HIJING}.
A number of sources of systematic error were studied, including
nuclear radius, nuclear skin depth, nucleon-nucleon inelastic
cross-section $\sigma_{\rm NN}$, and minimum nucleon separation.
The systematic error contributed by each source was determined
by varying that specific parameter in the analysis within reasonable
limits and quantifying the change in the final eccentricity result
as a function of centrality. The individual contributions were added
in quadrature to determine the 90\% confidence level errors shown
in Figure~\ref{aveecc}.

The crucial importance of the definition of eccentricity in
comparing Cu-Cu and Au-Au results can be seen in
Figure~\ref{v2nparteccscale}, where comparisons are made between
Cu-Cu and Au-Au data using the eccentricity-scaled elliptic flow.
In Fig.~\ref{v2nparteccscale}b, v$_2/\varepsilon_{std}$ increases rapidly
in Cu-Cu as the events become more central, and is generally larger than
that of Au-Au.  One might then conclude from this that either the smaller
Cu-Cu system produces v$_2$ much more efficiently than the larger Au-Au system
or that $\varepsilon_{std}$ may not be the appropriate quantity for describing
the initial geometry of the collision.  Consider then Fig.~%
\ref{v2nparteccscale}c, in which v$_2/\varepsilon_{part}$ is shown to be
very similar for both Cu-Cu and Au-Au, even appearing to lie on the same curve.
Given the qualitative and quantitative similarities between the
results in the two systems already shown, it is not unreasonable to expect
both systems to have a similar eccentricity-scaled elliptic flow, as in Fig.~%
\ref{v2nparteccscale}c.  Therefore, 
it seems likely that $\varepsilon_{part}$ as discussed here and in
Ref.~\cite{ManlyQM05} --- or a rather similar quantity, such as
$\sqrt{\langle \varepsilon^2_{part} \rangle}$~\cite{Bhal06} ---
is the relevant eccentricity for the azimuthal anisotropy.

In summary, the results presented here show a measurable and
significant elliptic flow signal in Cu-Cu collisions at 62.4 and 200 GeV. 
These data show that qualitative
features attributed to collective effects in Au-Au persist down to
the relatively small numbers of participants seen in the Cu-Cu
collision and are of comparable magnitude.  The essential role of
the choice of collision geometry in comparing flow across nuclear
species is clearly demonstrated.  If it is assumed that the flow
is independent of species for a given collision geometry, the data
shown here strongly suggest that it is the participant eccentricity,
not the nuclear eccentricity, that is responsible for elliptic flow.
These results
also imply that $\Psi_2$ is an estimate of $\Psi_{\rm participant}$,
which may be the relevant angle, particularly for systems with smaller
numbers of participants. This,
in turn, may provide information on the nature of the
matter driving the flow.

{\bf Acknowledgements:}
%
%
%
%
This work was partially supported by U.S. DOE grants
DE-AC02-98CH10886,
DE-FG02-93ER40802,
DE-FC02-94ER40818,  
DE-FG02-94ER40865,
DE-FG02-99ER41099, and
W-31-109-ENG-38, by U.S.
NSF grants 9603486, 
0072204,            
and 0245011,        
by Polish KBN grant 1-P03B-062-27(2004-2007),
by NSC of Taiwan Contract NSC 89-2112-M-008-024, and
by Hungarian OTKA grant (F 049823).


\end{document}